\documentclass[conference]{IEEEtran}

\usepackage{color}
\usepackage[english]{babel}
\usepackage{booktabs}
\usepackage{paralist}
\usepackage{graphicx}
\usepackage{multirow}
\usepackage{soul}
\usepackage{amsmath}
\usepackage{cleveref}
\usepackage{url}
\usepackage{float}
\usepackage{cite}
\usepackage{enumerate}
\usepackage{enumitem}
\usepackage{fixltx2e}
\usepackage{algorithm}
\usepackage[noend]{algpseudocode}
\usepackage{relsize}
\usepackage{subfig}
\usepackage{texdef2015}
\usepackage{setspace}
\usepackage{balance}
\usepackage{epstopdf}
\newcommand{\age}{\Delta}

\usepackage{amstext} 
\usepackage{array}   
\newcolumntype{L}{>{$}l<{$}} 

\usepackage{cleveref}
\crefname{section}{§}{§§}
\Crefname{section}{§}{§§}

\newcommand{\RY}[1]{}

\def\BibTeX{{\rm B\kern-.05em{\sc i\kern-.025em b}\kern-.08em
    T\kern-.1667em\lower.7ex\hbox{E}\kern-.125emX}}

\begin{document}

\title{An Empirical Study of Ageing in the Cloud}

\author{\IEEEauthorblockN{Tanya Shreedhar}
\IEEEauthorblockA{\textit{Wireless Systems Lab, IIIT-Delhi} \\
tanyas@iiitd.ac.in}
\and
\IEEEauthorblockN{Sanjit K. Kaul}
\IEEEauthorblockA{\textit{Wireless Systems Lab, IIIT-Delhi} \\
skkaul@iiitd.ac.in}
\and
\IEEEauthorblockN{Roy D. Yates}
\IEEEauthorblockA{\textit{WINLAB, Rutgers University} \\
ryates@winlab.rutgers.edu}
}

%
\maketitle
%
\IEEEpubidadjcol




\begin{abstract}
We quantify, over inter-continental paths, the ageing of TCP packets, throughput and delay for different TCP congestion control algorithms containing a mix of loss-based, delay-based and hybrid congestion control algorithms.
In comparing these TCP variants to  ACP+, an improvement over ACP, we shed better light on the ability of ACP+ to deliver timely updates over fat pipes and long paths.
ACP+ estimates the network conditions on the end-to-end path and adapts the rate of status updates to minimize age. It achieves similar average age as the \textit{best} (age wise) performing TCP algorithm but at end-to-end throughputs that are two orders of magnitude smaller. We also quantify the significant improvements that ACP+ brings to age control over a shared multiaccess channel.

\end{abstract}  


\section{Introduction}
\label{sec:introduction}
The challenge of age control over an end-to-end path in the cloud
is to adapt the rate of status updates entering the path so that
there are as many status updates in transit as possible while no update waits behind another in a router queue.
This is in contrast to Transmission Control Protocol (TCP) loss-based congestion control algorithms that aim for high throughput by having as many packets as possible in each queue without exceeding the available queue occupancy. This allows an end-to-end flow to achieve a rate equal to the bottleneck bandwidth; however, this is at the expense of large delays and eventual losses due to excessive queueing at the bottleneck.

Recently, requirements of low latency have led to the proposal of hybrid congestion control mechanisms such as BBR \cite{cardwell-bbr-2016}. At its stated ideal point of operation, BBR would have TCP packets delivered to the receiver at the bottleneck link rate, while each packet would experience an average delay as that experienced by a packet if only it was sent over the path. Intriguingly, this would satisfy the goal of age control by resulting in the highest rate of packet delivery at the receiver and lowest packet delays.

In our prior work \cite{tanya-kaul-yates-wowmom2019}, we had proposed the Age Control Protocol (ACP),  a transport layer protocol that regulates the rate at which status updates are sent by an application over an end-to-end path. By abstracting away an end-to-end path as a series of queues, we had argued that a good age control algorithm must try to have as many status updates in transit as possible while trying to ensure that the updates don't  wait for  previously queued prior updates from the application. 

\begin{figure}[t]             
	\begin{center}
	        \subfloat[\small Update rate high, delay high, age high]{\includegraphics[width=.45\textwidth]{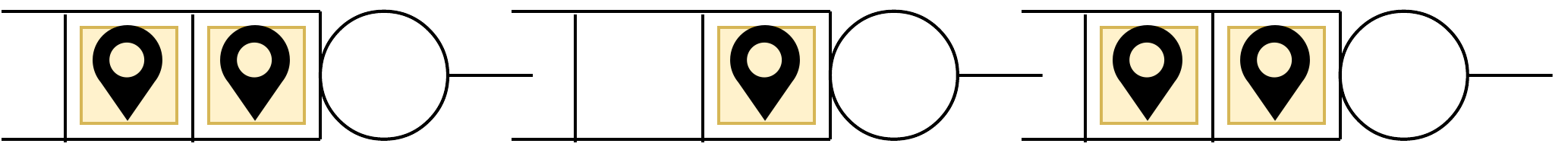}
			\label{fig:goodACP_01}}
			\\
			\subfloat[\small Update rate low, delay low, age high]{\includegraphics[width=.45\textwidth]{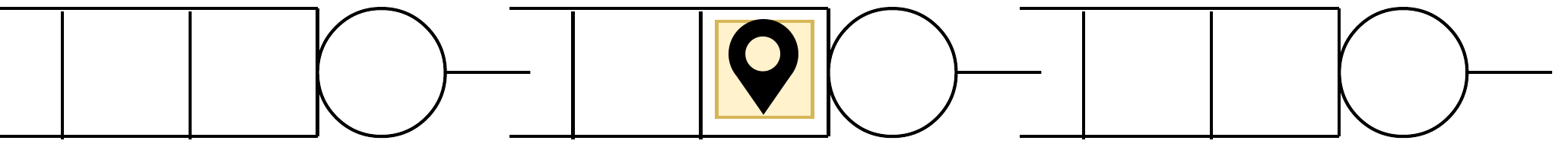}
			\label{fig:goodACP_02}}
			\\
			\subfloat[\small Ideal snapshot of updates in transit]{\includegraphics[width=.45\textwidth]{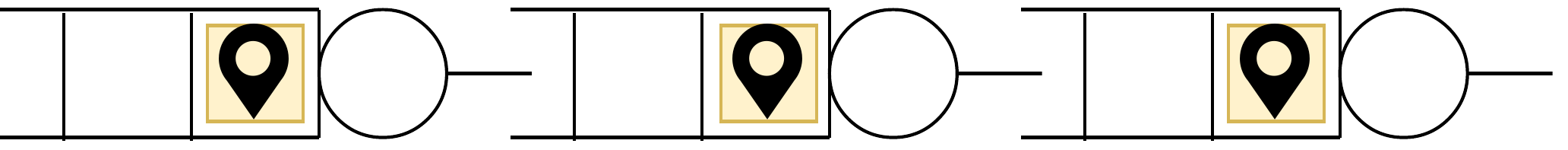}
			\label{fig:goodACP_03}}
		\caption{\small An illustration of queue occupancy and its impact on age.}
		\label{fig:goodACP}
	\end{center}
\end{figure}

Figure~\ref{fig:goodACP} provides an illustration of a good age control strategy in action for an end-to-end path of three identical queues, each  with deterministic service  times. Figures~\ref{fig:goodACP_01} and~\ref{fig:goodACP_02}, respectively, have too many and too few updates, resulting in high age. Figure~\ref{fig:goodACP_03} shows the snapshot one would expect to see with a good age control algorithm sending updates over the three-queue network. Of course, the picture becomes more complicated when the queues have random service times. For example, with a pair of M/M/1 queues in tandem, the average number of packets queued in the system at minimum age was shown to be $\approx 1.6$ updates~\cite{tanya-kaul-yates-arxiv}.

\begin{figure}[t]             
	\begin{center}
	        \subfloat[\small RTT as a function of offered load]{\includegraphics[width=.48\textwidth]{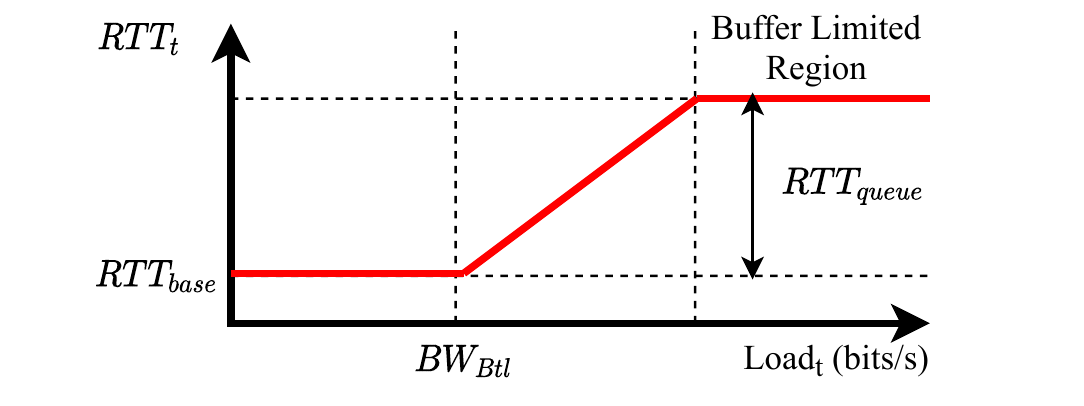}
			\label{fig:transient_tcp}}
			\\
			\subfloat[\small Average steady state RTT as a function of average load]{\includegraphics[width=.48\textwidth]{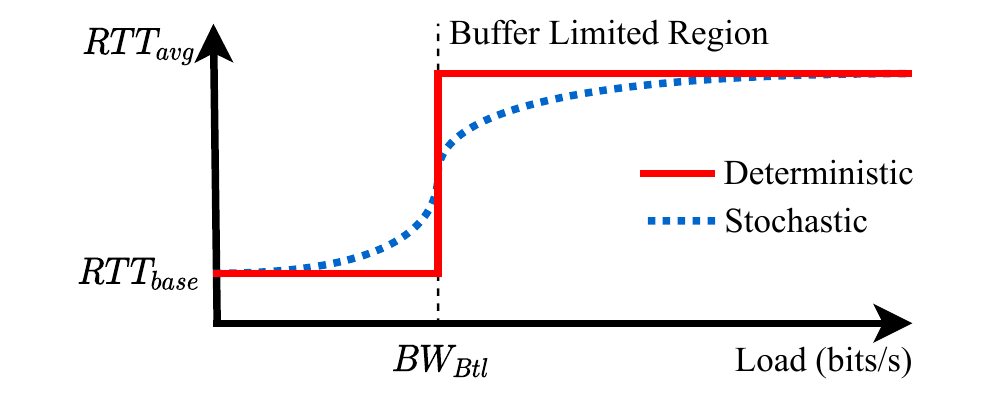}
			\label{fig:steady_tcp}}
		\caption{\small An illustration of how round-trip times vary as a function of the offered load. While (a) shows the change in instantaneous RTT as the load increases, (b) shows  the steady-state average behavior at a chosen load.\RY{This caption needs expanding to explain the difference between 2a and 2b}}
		\label{fig:TCPtransient_SteadyState}
	\end{center}
\vspace{-0.2in}

\end{figure}
While applications have diversified significantly over the past few decades, TCP congestion control remains the primary mechanism by which end hosts share the available bandwidth. For the purposes of TCP, the end-to-end path  may be abstracted away as a link with bottleneck bandwidth $BW_{\text{Btl}}$ and a round-trip propagation time of $RTT_{\text{base}}$~\cite{cardwell-bbr-2016}. Figure~\ref{fig:transient_tcp} provides an illustration, akin to that in~\cite[Figure $1$]{cardwell-bbr-2016}, of the instantaneous round-trip time $RTT_t$ at time $t$
as a function of the current offered load (the effective rate at which TCP is sending bytes).  As long as the offered load is smaller than $BW_{\text{Btl}}$, the TCP packets see a low RTT of $RTT_{\text{base}}$. Once the offered load becomes larger than $BW_{\text{Btl}}$, the TCP packets that arrive at the link's queue see increasingly more packets waiting for service ahead of them. This results in a linear increase in $RTT_t$ until 
the queue becomes buffer limited, the RTT saturates and TCP packets arriving at a full queue are dropped. 

Traditionally, TCP's congestion control allows for an increasing number of unacknowledged bytes from an application to flow through the network pipe until one or more bytes are lost due to the resulting congestion. Such a loss-based congestion control algorithm keeps increasing the offered load until a packet is lost as a result of the link operating in the buffer limited region. The flow will achieve a throughput equal to the bottleneck bandwidth, but packets in the flow will suffer large round-trip times, especially when the link has a large buffer.

Figure~\ref{fig:transient_tcp} suggests that one would like to operate at the lower ``knee'' in the curve, i.e., close to the bottleneck throughput $BW_{\text{Btl}}$ at low delays. In fact, delay-based and hybrid congestion control algorithms such as the recently proposed \emph{Bottleneck Bandwidth and Round-trip propagation time} (BBR) protocol, attempt this by using the round-trip time to detect congestion early, before a loss occurs due to buffer unavailability at a certain router along the path. 
Note that this combination of a throughput of $BW_{\text{Btl}}$ and round-trip times of $RTT_{\text{base}}$ is in fact achieved by the snapshot in Figure~\ref{fig:goodACP_03} that illustrates a good age control algorithm in action.

Of course, as was observed in~\cite{kleinrock2018internet} in relation to the stated point of operation of BBR, when a path is better modeled by a stochastic service facility, the average round-trip times at the maximum achievable throughput of $BW_{\text{Btl}}$ could be much larger than $RTT_{\text{base}}$. Figure~\ref{fig:steady_tcp} provides an illustration of steady-state average RTT as a function of average load. The red and blue curves, respectively, correspond to a deterministic and a stochastic service facility.

\begin{figure}[!t]             
\begin{center}
\includegraphics[width=0.9\linewidth]{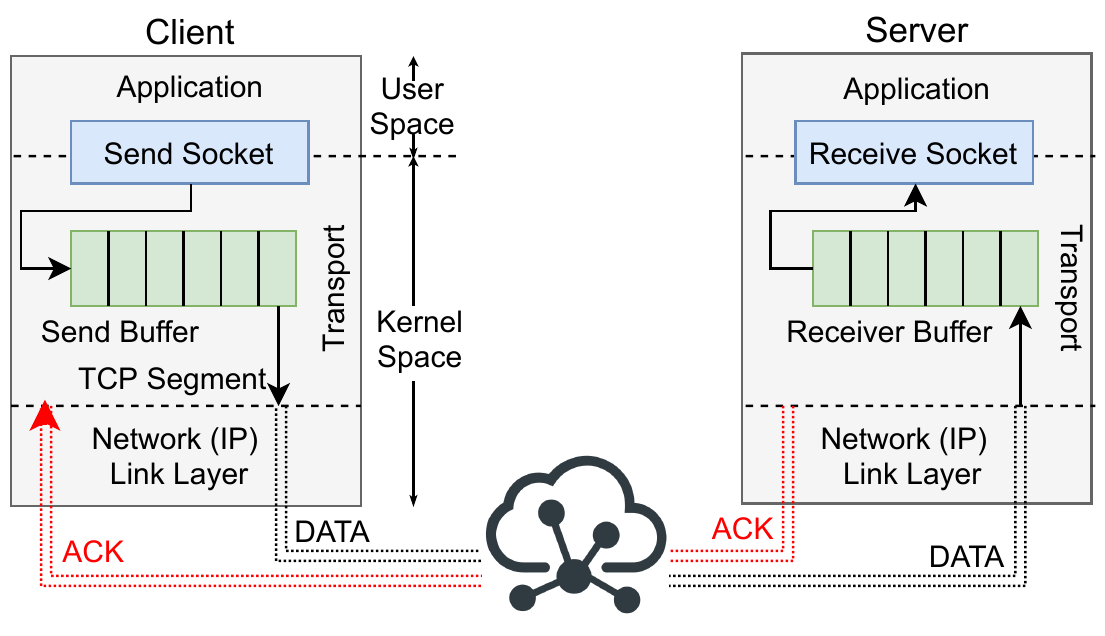}
\caption{\small TCP Network Stack}
\label{fig:TCP-Stack}
\end{center}
\vspace{-0.2in}
\end{figure}

This shift in congestion control algorithms from keeping the pipe full to ``keeping the pipe just full, but no fuller''~\cite{kleinrock2018internet}, motivates this empirical study of how the information at a receiver would age if updates were transmitted over the cloud using the congestion control algorithms. However, we must be careful as (a) TCP doesn't regulate the rate of generation of packets by the status updating application, (b) it is a stream-based protocol and has no notion of update packets. As illustrated in Figure~\ref{fig:TCP-Stack}, an application  writes a stream of bytes to the TCP sender's buffer. TCP creates segments from these bytes in a first-come-first-serve manner. TCP segments are delivered to the TCP receiver. At any time, TCP allows a total of up to a current congestion window size of bytes to be in transit in the network. The TCP receiver sends an \texttt{ACK} to inform the sender of the last segment received.

To stay focused on evaluating how 
scheduling TCP segments over an end-to-end path would age updates at a receiver, we assume that a TCP segment, when created, contains fresh information. Specifically, we ignore the ageing of bytes while they wait in the TCP send buffer. One way of achieving this in practice would be to have the application provide freshly generated information (as in a generate-at-will model~\cite{yates2020age-survey}) to be incorporated in a TCP segment just as TCP schedules it for sending.
\RY{Is this possible with traditional layering?} 

Here, we approximate the age of the segment when it arrives at the TCP receiver to be the RTT of the segment, which is calculated based on the time of receipt of the TCP \texttt{ACK} that acknowledges receipt of the segment. Further, we approximate the inter-delivery time of segments at the receiver by the inter-delivery times of the corresponding \texttt{ACK}s. The RTT(s) and the inter-delivery times together allow us to come up with an estimate of the time-average of age at the receiver that results from a chosen congestion control algorithm, using the graphical method of time-average age calculation~\cite{KaulYatesGruteser-Infocom2012}.

Last but not the least, we would like to minimize the impact of packet loss due to link transmission errors on our evaluation of congestion control. Given our focus on paths in the cloud, specifically between AWS data centers, we observe a very small percentage of loss, and that too because of buffer overflows in routers that result in the process of congestion control estimating the bottleneck bandwidth.

Our specific contributions include:
\begin{enumerate}[leftmargin=*]
    \item We propose the ACP+ algorithm and explain the changes in it with respect to ACP. (~\cref{sec:algorithm_new})
    \item We provide an empirical study of age, throughput and delay trade-offs obtained when using state-of-the-art TCP congestion control algorithms to transport updates over an end-to-end path in the cloud. 
    
    We evaluate a mix of loss-based (Reno~\cite{reno} and CUBIC~\cite{ha2008cubic}), delay-based (Vegas~\cite{brakmo1994tcp}) and hybrid congestion control algorithms (YeAH~\cite{baiocchi2007yeah} and BBR~\cite{cardwell-bbr-2016}) for different settings of receiver buffer size.
    We compare the performance of the TCP algorithms with that of ACP+. We show that ACP+ does well to estimate the network conditions on the end-to-end path and appropriately adapts the rate of status updates sent over the path to keep age at the receiver close to the minimum. (~\cref{sec:real-world})
    \item We show that ACP+ provides significant improvements for an end-to-end path that includes a multiaccess hop shared by many status updating applications using \texttt{ns3} simulations. (~\cref{sec:evaluation})
\end{enumerate}

\noindent \textbf{Related Work:} The past five years have seen work on various aspects of the age of information; recent surveys can be found in~\cite{kosta2017-FTN}, ~\cite{yates2020age-survey}. That said, there is limited systems research \cite{Sonmez-BaghaeeSeaCom2018,shreedhar2018acp,tanya-kaul-yates-wowmom2019,kadota2020wifresh} on ageing of information and its optimization in real-world networks. In \cite{Sonmez-BaghaeeSeaCom2018}, authors discuss the age of information (AoI) in real-networks where a source is sending updates to a monitor over different access networks. The key takeaway from that work is the need for an AoI optimizer that can adapt to changing network topologies and delays. The Age Control Protocol (ACP) was proposed in~\cite{shreedhar2018acp,tanya-kaul-yates-wowmom2019}. ACP is a transport-layer solution that works in an application-independent and network-transparent manner. ACP attempts to minimize the age of information of a source update at a monitor connected over an end-to-end path on the Internet. Very recently, in~\cite{kadota2020wifresh} the authors proposed under the name of WiFresh a MAC-layer and an application-layer solution to ageing of updates over a wireless network. While both \cite{tanya-kaul-yates-wowmom2019} and \cite{kadota2020wifresh} look at ageing of updates on the Internet, they differ in their approach and scope. ACP is a transport layer solution that works by adapting the source generation rate without any specific knowledge of the access network or any network hop to the monitor, whereas, WiFresh is a scheduling solution designed for WiFi networks.


\section{The ACP+ Control Algorithm}
\label{sec:algorithm_new}
ACP+ is an improved version of ACP~\cite{tanya-kaul-yates-wowmom2019}. Just like ACP, it is a transport layer protocol that operates on the end-hosts. It uses UDP as a substrate and benefits from its unreliability feature to send updates from a source to a monitor over the Internet. Since ACP+ is quite similar to ACP, we provide a succinct summary of the algorithm and highlight the salient differences between ACP+ and ACP\footnote{The design principles and details about ACP can be found in ~\cite{tanya-kaul-yates-wowmom2019,tanya-kaul-yates-arxiv}.}. 

The  ACP+ source appends a header to the update containing a \emph{timestamp} field that stores the time it was generated. ACP+ suggests to the source the generation rate of updates. For this, it must estimate network conditions over the end-to-end path to the monitor. This estimation is enabled by having the  ACP+ monitor send back an \texttt{ACK} to the ACP+ source for every received update. The \texttt{ACK} contains the timestamp of the update it is acknowledging. 

\begin{algorithm}[t]
\caption{ACP+ Control Algorithm}
\label{alg:acp+}
\footnotesize
\begin{algorithmic}[1]
\State \textbf{INPUT:} $b_k, \delta_k, \overline{T}$, $B_k$
\State \textbf{INIT:} $flag \gets 0$, $\gamma \gets 0$
\While{true} 
	\If {$b_k>0$  \&\& $\delta_k >0$}\label{alg:one}
		 \If {$flag==1$}
		 \State $\gamma=\gamma+1$\label{alg:oneincr}
			\State MDEC($\gamma$): $b^{*}_{k+1} = -(1 - 2^{-\gamma}) B_k$
		\Else
			\State DEC: $b^{*}_{k+1} = -1$ \label{alg:oneDEC}
		\EndIf
		\State $flag\gets 1$
	\ElsIf { $b_k>0$  \&\& $\delta_k <0$}\label{alg:two}
	        \State INC: $b^{*}_{k+1} = 1$
			\State $flag\gets 0$,
			$\gamma\gets0$ \label{alg:two2}
	\ElsIf { $b_k<0$  \&\& $\delta_k >0$}\label{alg:three}
	    \State INC: $b^{*}_{k+1} = 1$
		\State $flag\gets 0$, $\gamma\gets0$ 
	\ElsIf { $b_k<0$  \&\& $\delta_k <0$}\label{alg:four}
		 \If {$flag==1$ \&\& $\gamma>0$}
			\State MDEC($\gamma$): $b^{*}_{k+1} = -(1 - 2^{-\gamma}) B_k$ \label{alg:fourMDEC}
		\Else
		    \State DEC: $b^{*}_{k+1} = -1$
			\State $flag\gets 0$, $\gamma\gets0$
		\EndIf
	\EndIf
\State \Call{UpdateLambda}{$b^{*}_{k+1}$} \label{alg:update}
\State wait $\overline{T}$ 	
\EndWhile
\Statex
\Function{UpdateLambda}{$b^{*}_{k+1}$}
\State $\lambda_k = \frac{1}{\overline{Z}} + \frac{b^{*}_{k+1}}{\mathcal{\overline{RTT}}}$ \label{alg:calc}
	\If {$\lambda_k < 0.75*\lambda_{k-1} $}\label{alg:min}
		 \State 
		 $\lambda_k = 0.75*\lambda_{k-1}$
		 \Comment{Minimum $\lambda$ threshold}
	\ElsIf {$\lambda_k > 1.25*\lambda_{k-1} $}\label{alg:max}
	        \State
	        $\lambda_k = 1.25*\lambda_{k-1} $
			\Comment{Maximum $\lambda$ threshold}
	\EndIf
\State
\Return $\lambda_k$
\EndFunction
\end{algorithmic}
\end{algorithm}

Consistent with AoI freshness metrics, ACP+ discards an \emph{out-of-sequence} packet  at the monitor and an \emph{out-of-sequence} \texttt{ACK} at the source. 
Algorithm \ref{alg:acp+} details the ACP+ control algorithm that is used to set the source's update rate $\lambda_k$ at time $t_k$.
The control algorithm executing at time $t_k$, uses an estimate of the time average update age ($\overline{\age}_k$) at the monitor and the time average of backlog ($\overline{B}_k$) calculated over the interval $(t_{k-1}, t_k)$, where $k$ indexes the current control epoch.

ACP+ uses RTT(s) of updates for age estimation and maintains an exponentially weighted moving average (EWMA) $\overline{\text{RTT}}$ of measured RTT(s). RTT is calculated for every update whose ACK is received. It is the time between the generation of the update and reception of the corresponding ACK. The ACP+ source also keeps an estimate $\overline{Z}$ of the EWMA of the time elapsed at the monitor between reception of consecutive updates. This time between reception of consecutive updates is approximated by the source as the time elapsed between its reception of the corresponding ACK(s). As seen in line~\ref{alg:calc} of Algorithm~\ref{alg:acp+}, and explained later, $\overline{\text{RTT}}$ and $\overline{Z}$ are used to calculate the source update rate $\lambda_k$. 
\RY{What is $\overline{Z}$ and how is it different from $\overline{\text{RTT}}$ Is it the EMA RTT estimate at the monitor?? Text doesn't seem to say how $\overline{Z}$ is used.}

The length $\overline{T}$ of a control epoch is set as $\mathcal{T} = 10/\lambda_{k}$. This ensures at least $10$ packets are sent by the source using the updated $\lambda_{k}$.
The source updates $\overline{\text{RTT}}$, $\overline{Z}$ and $\overline{T}$ every time an \texttt{ACK} is received.
At every control epoch $k > 1$, at time $t_k$, the ACP+ source calculates the differences $\delta_k = \overline{\age}_k - \overline{\age}_{k-1}$ and $b_k = \overline{B}_k - \overline{B}_{k-1}$.

At the source, ACP+  chooses an action $u_k$ at the $k$\textsuperscript{th} epoch that targets a change $b^{*}_{k+1}$ in average backlog over an interval of length $\mathcal{T}$ with respect to the $k$\textsuperscript{th} interval.
The actions, may be broadly classified into (i) additive increase \texttt{(INC)}, (ii) additive decrease \texttt{(DEC)} and (iii) multiplicative decrease \texttt{(MDEC)}. \texttt{MDEC} corresponds to a set of actions \texttt{MDEC}$(\gamma)$, where $\gamma = \{1,2,\ldots\}$.

The  ACP+ source targets a reduction in average backlog over the next control interval in case either \{$b_k>0$, $\delta_k>0$\} (line~\ref{alg:one} in the algorithm) or \{$b_k<0$, $\delta_k<0$\} (line~\ref{alg:four}).
The ACP+ source targets an increase in average backlog over the next control interval in case either \{$b_k>0$, $\delta_k<0$\} (line~\ref{alg:two}) or \{$b_k<0$, $\delta_k>0$\} (line~\ref{alg:three}). 
Algorithm~\ref{alg:acp+} summarizes how ACP+ chooses its action $u_k$ as a function of $b_k$ and $\delta_k$ to achieve the desired $b^{*}_{k+1}$. Next, ACP+ calls the function \texttt{UPDATELAMBDA} (line~\ref{alg:update}) that sets the rate at which the source must send updates to achieve the target backlog $b^{*}_{k+1}$ (line~\ref{alg:calc}).
However, to restrict the unnecessary oscillations  close to the age minimizing $\lambda$, we use \texttt{min} 
and \texttt{max} thresholds  (lines~\ref{alg:min}-\ref{alg:max}) to restrict the range of $\lambda_{k}$.


The most significant change in ACP+ over ACP is in the function \texttt{UPDATELAMBDA}. ACP used a step size parameter $\kappa$ in \texttt{(INC)} and \texttt{(DEC)}. Instead of $\pm 1$, the desired change in backlog was set to $\pm \kappa$; however, we found $\kappa$ was remarkably difficult to set. For example, while in our real experiments with ACP $\kappa=1$ worked well, in simulations with shorter paths, with small round-trip times but not so small propagation times, age control went haywire. For the latter, a small value of $\kappa = 0.25$ ensured proper updating of the update rate $\lambda_k$.
\RY{Are we referring to $\lambda_k$ as the ACP rate?} The $\kappa$ based update could also result in $\lambda_k$ (calculated in line~\ref{alg:calc}) becoming very small or even negative. To avoid this, we clamped the minimum $\lambda_k$ to be at least one packet per round-trip time. Unfortunately, this resulted in high age in settings where multiple ACP paths shared a constrained access. ACP+ doesn't use such a minimum. Instead \texttt{UPDATELAMBDA} restricts the step change in $\lambda$. This, as shown in \cref{sec:evaluation}, results in significant improvements in age achieved when a large number of ACP+ sources send updates over a shared multiaccess.

\section{Real-World Setup and Experiments}
\label{sec:real-world}
\begin{figure}[!t]             
\begin{center}
\includegraphics[width=1\linewidth]{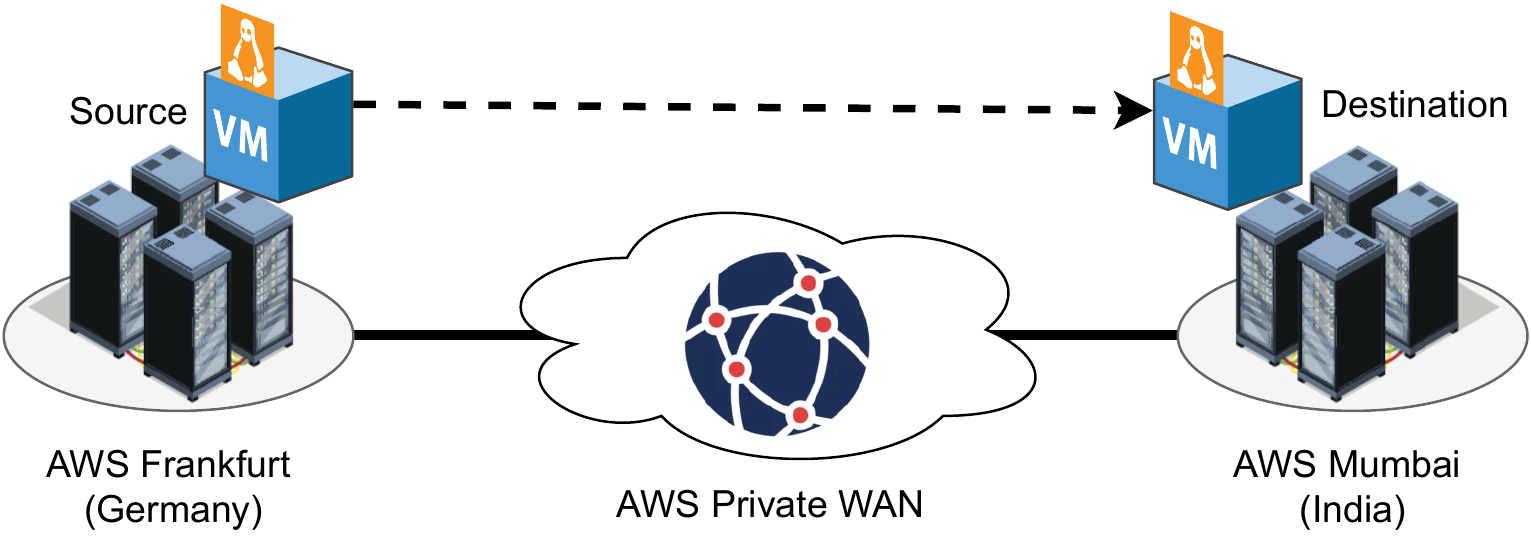}
\caption{\small An illustration of the real experiment topology on the AWS E2 cloud network. The client machine (both ACP+/TCP) was in AWS Frankfurt, Germany, and the server was in AWS Mumbai, India. The instances were connected via the AWS Private WAN.}
\label{fig:AWS-setup}
\end{center}
\vspace{-0.2in}
\end{figure}

In this section, we describe the experimental setup and methodology, followed by an empirical comparison of ACP+ and the chosen congestion control algorithms with respect to the metrics of age, throughput and delay. The ACP+ client and server codes are available at \cite{acpGitRepo}.

\textbf{Methodology:} Figure~\ref{fig:AWS-setup} shows the real experiment topology. All our experiments over the Internet used two T2.micro instances in the AWS EC2 cloud network. Both instances are configured with one virtual CPU, $1$ GB RAM and a $1$ Gbps Ethernet link connected to the AWS private WAN. One of the instances was in the AWS Frankfurt (Germany) data-center, while the other was deployed in the AWS Mumbai (India) data-center. Each instance ran a virtual machine with Ubuntu $18.04$ LTS with Linux kernel version $5.3$. We confirmed through periodic \texttt{traceroute} that the underlying network between our two chosen instances was served by the AWS private WAN. 

We describe our measurement methodology next. For both the ACP+ and TCP experiments, we deployed the sender in AWS Frankfurt and the receiver in AWS Mumbai. 
For each chosen congestion control algorithm, we investigated the impact of different receive buffer sizes on the performance of the congestion control algorithms by changing \texttt{default} and \texttt{maximum} values of \texttt{r\_mem} in the Linux kernel. The space available in the receiver buffer limits the maximum amount of bytes that any congestion control algorithm may send to the TCP receiver.

For the TCP experiments, we used \texttt{iPerf3} for packet generation and \texttt{Wireshark} for packet captures. To ensure that all algorithms saw similar network conditions we ran multiple iterations of ACP+, TCP BBR, TCP CUBIC, TCP Reno, TCP Vegas and TCP YeAH, in that exact order, one after the other. For each TCP variant in the stated order, we further ran different receive buffer settings. Each run of the experiment lasted $200$ s.
Considering that end-to-end RTT is $\approx 110$ ms in our setup, 
TCP spends a majority of the transfer time in the steady-state phase. 

\textbf{Results:} We show results from $40$ runs each of ACP+, BBR-d1m1, BBR-d1m3, BBR-d5m5, CUBIC, Reno, Vegas and YeAH. For each run, we show the average age, throughput, and average delay (round-trip time). In the above list, we have BBR run with three different receiver buffer settings. BBR-d1m1 denotes the smallest \texttt{default} and \texttt{maximum} values of the receiver buffer (\texttt{r\_mem}). In BBR-d1m3, the \texttt{default} is the same as BBR-d1m1 but the \texttt{maximum} is three times larger. Similarly, in BBR-d5m5 both the \texttt{default} and the \texttt{maximum} is five times that in BBR-d1m1. For all other TCP algorithms, the results are shown for a \texttt{default} and a \texttt{maximum} five times that of BBR-d1m1. In general, one would expect a larger receiver buffer to allow the TCP algorithm to have a larger number of bytes in flight as long as the network doesn't become the bottleneck.

\begin{figure}[!t]             
\begin{center}
\includegraphics[width=1\linewidth]{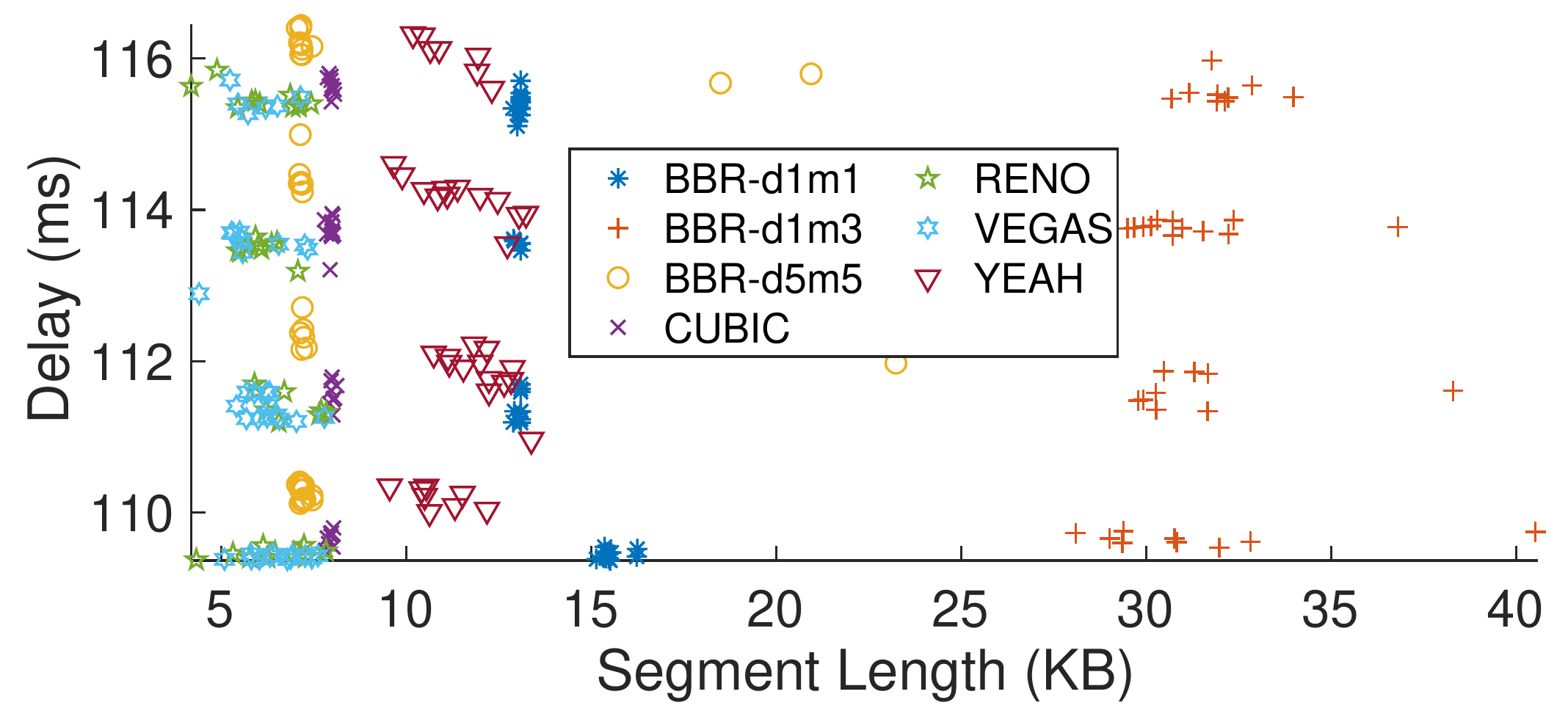}
\caption{\small TCP segment length vs. delay obtained for the runs of the different algorithms.}
\label{fig:delay_segLen_all}
\end{center}
\vspace{-0.2in}
\end{figure}

\textbf{Queue Waiting Delays Dominate:} Figure~\ref{fig:delay_segLen_all} shows the impact of TCP segment lengths on delay. As is seen, segment length and delays are uncorrelated 
\RY{Is the range of delays from 109ms to 114 ms as in Fig 5?? Referring to this tiny range as the  entire range od delays seems weird.} for all the TCP algorithms. This observation can be explained by the fact that the delays in the network are almost entirely because of the time spent in router queues awaiting transmission. The transmission times (propagation delays), which are about $20$ ms, are a small fraction in comparison.\RY{Propagation is on the order of 20-40ms between the AWS centers? Frankfurst to India must be as far as NY to LA?? If so, that's only a fraction of 110 but not negligible.} It may be worth noting that the TCP segment lengths are chosen by the TCP algorithm and often change during a TCP session. In the figure, we show segment lengths averaged over a run.

\begin{figure}[!t]             
\begin{center}
\includegraphics[width=1\linewidth]{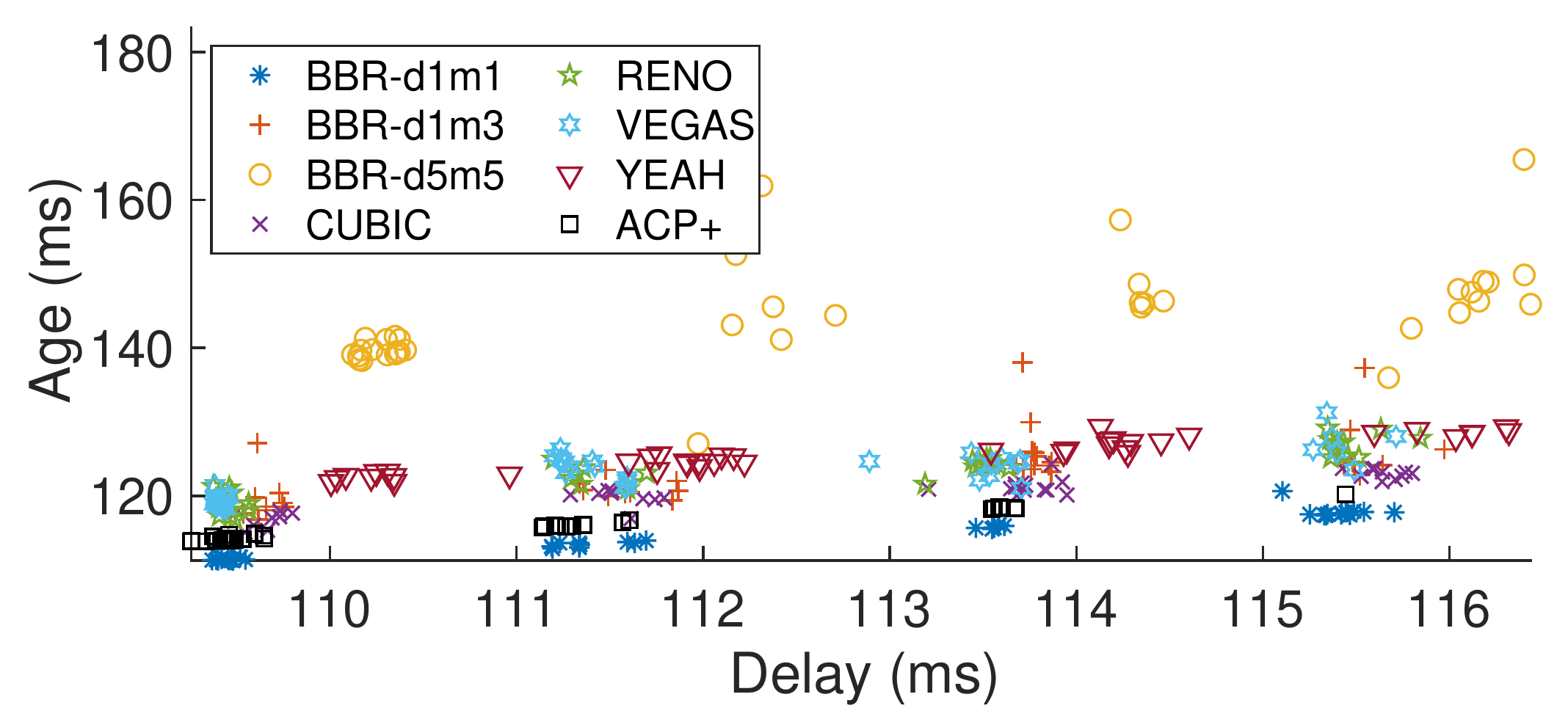}
\caption{\small Delay vs. age for the different runs of the chosen algorithms.}
\label{fig:delay_age_all}
\end{center}
\vspace{-0.2in}
\end{figure}

\textbf{Delay vs. Age:} Figure~\ref{fig:delay_age_all} shows a scatter of (delay, age) for the chosen runs. We see that BBR-d5m5 sees both age and delays larger than the rest. Amongst the rest, from the figure, it is apparent that ACP+ achieves delays and ages smaller than all algorithms other than BBR-d1m1. BBR-d1m1 achieves a slightly smaller age than ACP+. 

In fact, the age and delay achieved by BBR-d1m1, averaged over all runs, are $114.5$ ms and $112.33$ ms, respectively. The corresponding values for ACP+ are $115.5$ ms and $110.79$ ms. The next smallest age is achieved by CUBIC and is $\approx 121$ ms. Reno, Vegas and BBR-d1m3 achieve higher ages than CUBIC, with YeAH achieving the highest age of about $125$ ms among them. BBR-d1m4, BBR-d1m5 and BBR-d5m5 achieve ages larger than $140$ ms. Only BBR-d5m5 is shown.

\begin{figure}[!t]             
\begin{center}
\includegraphics[width=1\linewidth]{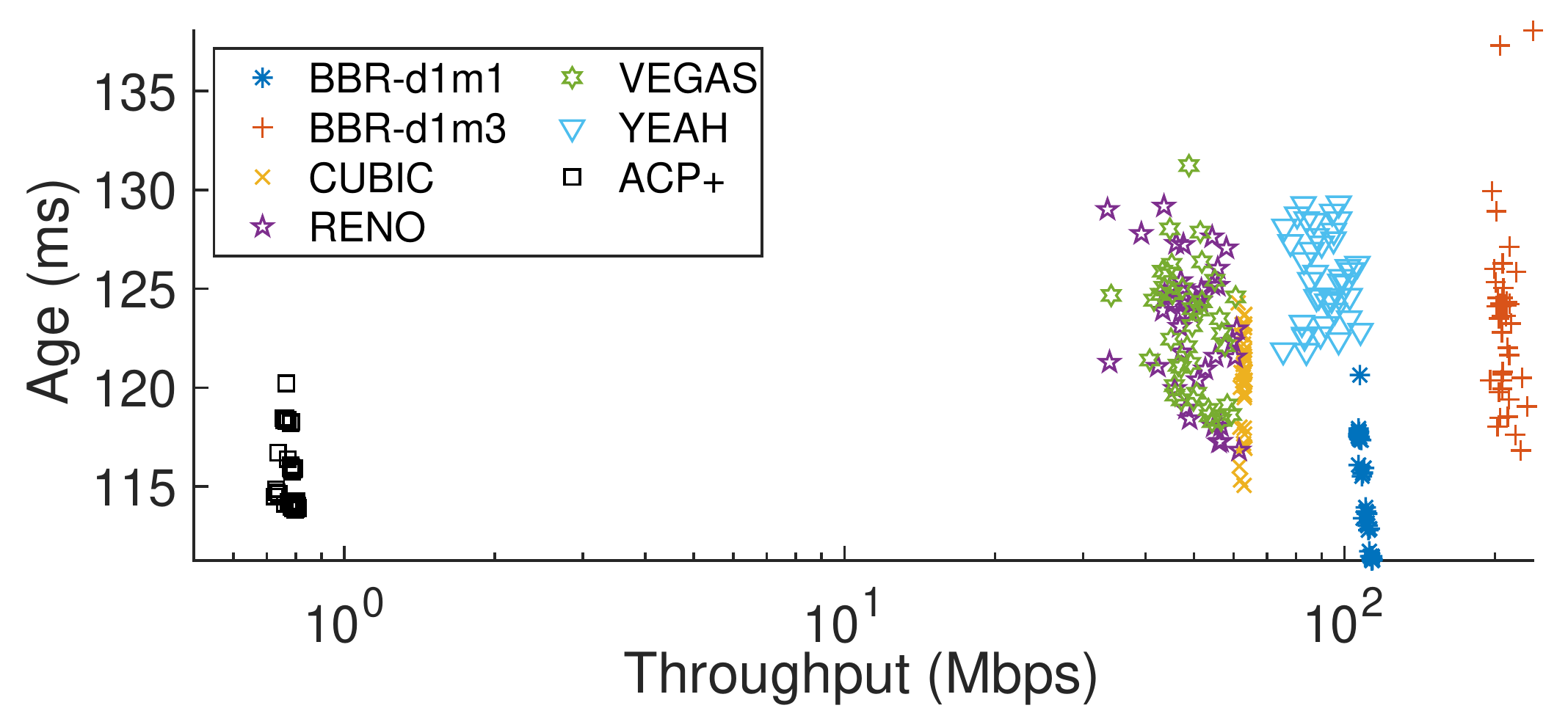}
\caption{\small Throughput vs. age for the different runs of the chosen algorithms.\RY{THIS IS AN AWESOME RESULT}}
\label{fig:thr_age_all}
\end{center}
\vspace{-0.2in}
\end{figure}

\textbf{ACP+ vs. BBR-d1m1:} Before we delve further into the relative performances of ACP+ and BBR-d1m1, let's consider Figure~\ref{fig:thr_age_all} in which we show the (throughput, age) values achieved by the different algorithms. We omit BBR-d5m5 from the figure as it resulted in high age values (average larger than $140$ ms) and also did not yield very good throughput. BBR-d1m3 achieves the highest throughput. In fact, its throughput of about $200$ Mbps is twice the next highest value of about $110$ Mbps achieved by BBR-d1m1. The average age when using BBR-d1m3 is $123.5$ ms in contrast to the $114.5$ ms obtained when using BBR-d1m1. 

Interestingly, the throughput obtained by ACP+ is a low of $0.77$ Mbps in contrast to $110$ Mbps obtained using BBR-d1m1 ($\approx 141 \times$ the ACP+ throughput). This stark difference is partly explained by the segment\footnote{Recall our assumption that every new segment contains a fresh update.} sizes used by BBR-d1m1, on an average about $14$ KB, in comparison to the constant $1024$ byte payload of an ACP+ packet. \RY{Why are the payloads so different? Somehow this should be related to a standard size update that is being used?} 
This difference still leaves an unexplained factor of about $10$. This is explained by an average inter-ACK time of $10.4$ ms for ACP+ in comparison to a much smaller $1.16$ ms for BBR-d1m1 that results from BBR-d1m1 attempting to achieve high throughputs. \RY{... because ...Why??}

To summarize, ACP+ results in an average age of $115.5$ ms, an average delay of $110.79$ ms, an average throughput of $0.77$ Mbps and an inter-ACK time of $10.4$ ms. The corresponding values for BBR-d1m1 are $114.5$ ms, $112.33$ ms, $110$ Mbps and $1.16$ ms. \emph{ACP+ achieves an almost similar age as BBR-d1m1, however, at a significantly lower throughput.} The similar age at a much larger inter-ACK time is explained by the fact (observed in our experiments) that while a very low or high rate of updates results in high age, age stays relatively flat in response to a large range of update rates in between.  It turns out that ACP+ tends to settle in the flat region closer to where increasing the rate of updates stops reducing age. This much reduced throughput of ACP+ is especially significant in the context of shared access, allowing a larger number of end-to-end ACP+ flows to share an access without it becoming a bottleneck. 

\textbf{The BBR Puzzle:} What could explain the low age achieved by BBR-d1m1? We observe that the average delay of $112.33$ ms when using BBR-d1m1 is the same as that obtained by a \emph{Lazy} (introduced in~\cite{tanya-kaul-yates-wowmom2019}) status updating protocol we ran alongside the others, which sends an update once every round-trip time. One would expect \emph{Lazy} to achieve a round-trip time of $RTT_{\text{base}}$ (see Figure~\ref{fig:transient_tcp}). This tells us that BBR-d1m1's flow on an average saw an RTT of $RTT_{\text{base}}$. While it obtained a low throughput of $100$ Mbps, it seems to have kept the pipe full enough. This low throughput was an accidental consequence of the receiver buffer size settings of BBR-d1m1, which disallowed the congestion control algorithm to push bytes into the network at a larger rate. The higher throughput achieved by BBR-d1m3, as observed earlier, came with a higher age, however.

\section{Simulations Setup and Results}
\label{sec:evaluation}
\begin{figure}[!t]             
\begin{center}
\includegraphics[width=\linewidth]{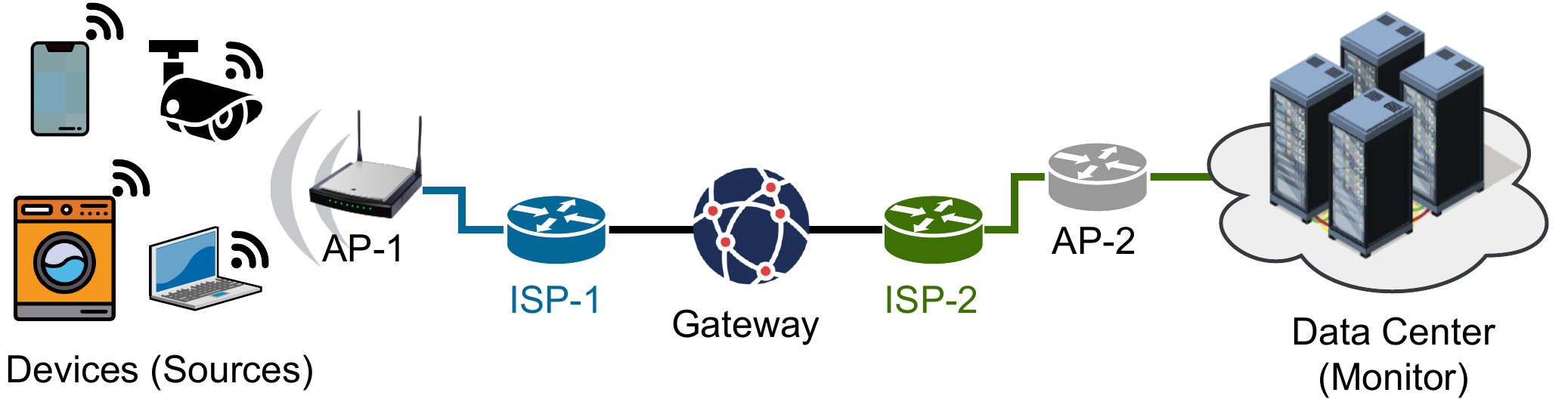}
\caption{\small Simulation Setup}
\label{fig:setup}
\end{center}
\vspace{-0.2in}
\end{figure}

We used the network simulator ns3\footnote{\url{https://www.nsnam.org/}} together with the YansWiFiPhyHelper\footnote{\url{https://www.nsnam.org/doxygen/classns3_1_1_yans_wifi_phy.html}}.
The base network topology used in our simulations is shown in Figure~\ref{fig:setup}. We show results for when source nodes are spread uniformly and randomly over an area of $20\times 20$ m$^2$. We chose the number of sources from the set $\{1,6,12,24,48\}$. The channel between the source and AP-1 was log-normally distributed with a standard deviation of $12$ and a path loss exponent of $3$. The WiFi link rate was set to $12$ Mbps and that of the P2P links was set to $6$ Mbps.

We compare the performance of ACP+ to \emph{Lazy}, which as mentioned before, is a conservative status updating mechanism that sends one update per $\overline{\text{RTT}}$ and maintains an average backlog of $1$ update in the network for a given source.

\begin{figure}[!t]             
	\begin{center}
		\subfloat[]{\includegraphics[width=.25\textwidth]{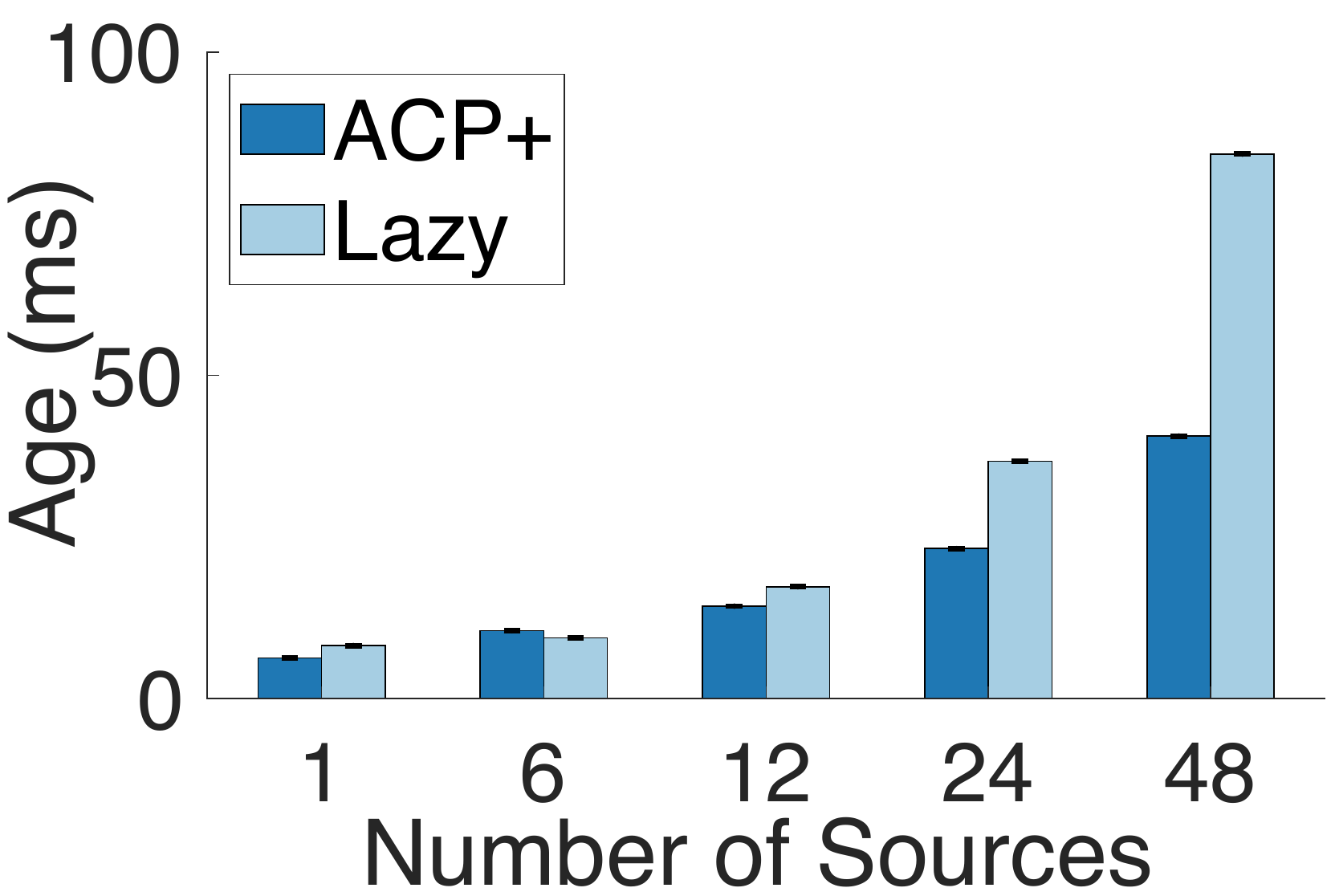}
			\label{fig:age_topology_1}}
		\subfloat[]{\includegraphics[width=.25\textwidth]{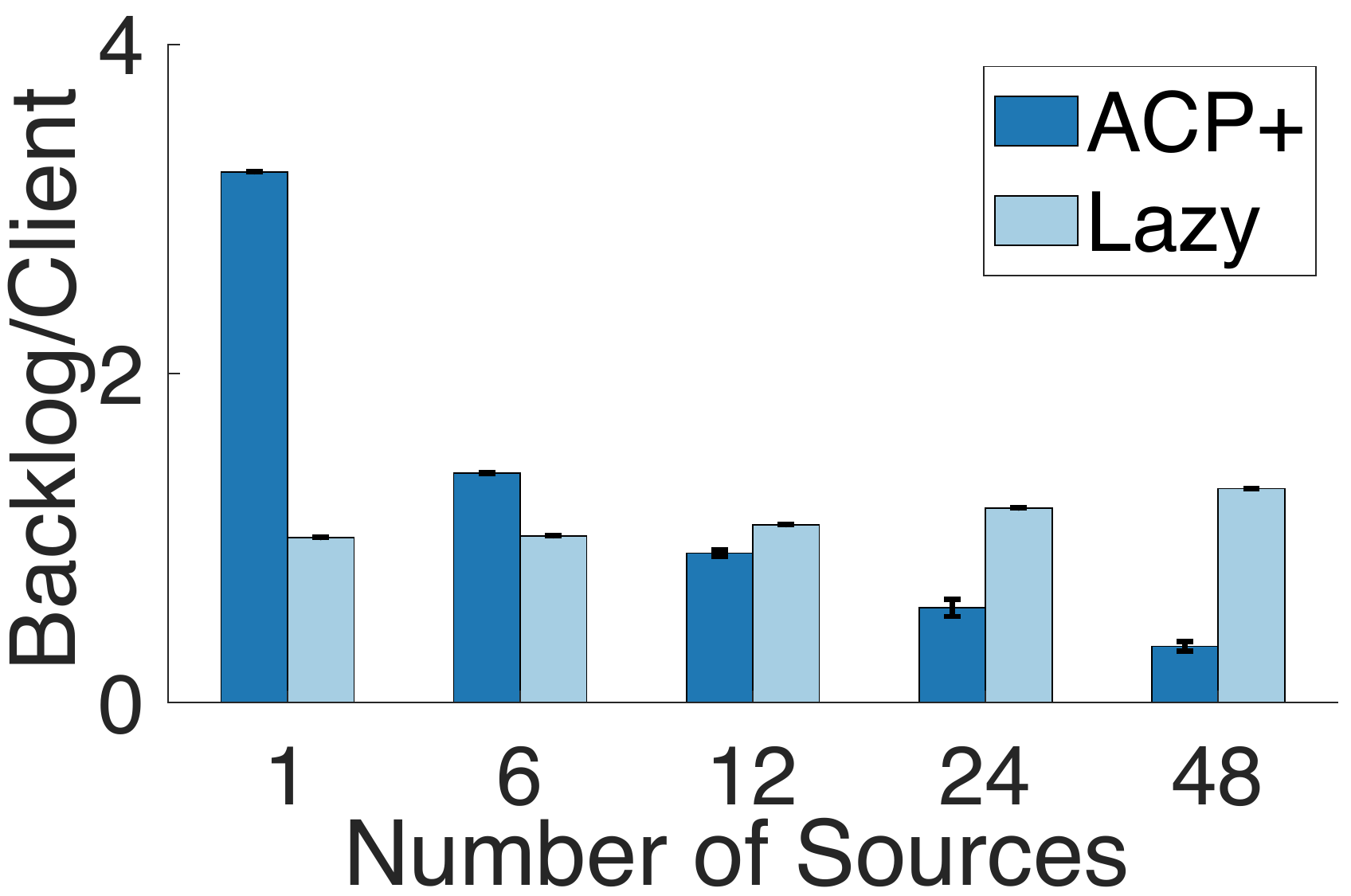}
			\label{fig:backlog_topology_1}}
        \caption{\small Average (a) source age and (b) source backlog}
		\label{fig:ageAndBacklogFastAndSlowNets}
	\end{center}
\vspace{-0.2in}
\end{figure}

Figure~\ref{fig:age_topology_1} shows that ACP+ achieves a smaller age per source than \emph{Lazy}. The improvements are especially significant when a large number of sources share the access to AP-1. That ACP+ is able to achieve smaller ages can be understood via the average backlog per source when using ACP+ and \emph{Lazy}, which is shown in Figure~\ref{fig:backlog_topology_1}. When we have just one source, ACP+ tries to fill each queue in the network with an update. This results in a larger backlog and a lower age in comparison to \emph{Lazy}, which achieves a backlog of just $1$ update. However, as the number of sources increases, while \emph{Lazy} continues to maintain a backlog of $1$ per source, ACP+ reduces it. The backlogs obtained are $3.23, 1.39, 0.91, 0.57, 0.34$, respectively, for $1,6,12,24,48$ sources.

The ACP+ backlogs when we have a large number of sources are not only much smaller than \emph{Lazy}, it turns out that they are not too far from an ideal scheduling mechanism that schedules updates from the sources in a round-robin manner. For simplicity, ignore the difference in the link rates of the WiFi and P2P links. Also, assume that no packets are dropped due to channel errors over WiFi. A round-robin scheduler would keep six updates in transit of the source when we have just one source. This would result in a backlog of $6$. It would schedule six sources one after the other in a manner such that a round of scheduling would lead to six packets in the six queues from the six different sources, resulting in an average backlog of $1$ per source. Similarly, for when we have $12, 24, 48$ sources, we would see backlogs per source of $1/2, 1/4, 1/8$, respectively. ACP+ sees larger backlogs than these, at least partly because of packet collisions over the WiFi access, which results in larger delays in the WiFi hop.

ACP+'s good adaptation to an increase in the number of sources is also seen in the fact that the RTT doesn't increase much as the number of sources increase. This is unlike \emph{Lazy} which sees big increases in RTT. While ACP+ results in RTT of $5.5, 7.1, 7.9, 9, 10.4$ ms, respectively, for $1,6,12,24,48$ sources, \emph{Lazy} sees RTT of $5.5, 6.3, 11.8, 26.3, 61.1$ ms.

ACP+ is \textit{age fair} with Jain's fairness index \cite{jain1984quantitative} reducing from about $.99$ to $.89$ as we increase source density from $6$ to $48$ nodes. As regards ACP, in such settings it had earlier been evaluated to be as bad as \emph{Lazy} at age control.

\section{Conclusions}
\label{sec:conclusions}
In this paper, we detail the design of ACP+, an improved version of the transport layer age control protocol ACP. We evaluated ACP+ and a mix of loss-based, delay-based and hybrid TCP congestion control algorithms over an inter-continental path connecting AWS data-centers in Mumbai, India and Frankfurt, Germany. We showed that ACP+ achieves a small age at a very low throughput. We also evaluated ACP+ in a setting where multiple sources send their updates using ACP+ over a shared multiaccess. We showed that the different ACP+ flows coexist well with each other and the sources see ages much lower than the old ACP and \emph{Lazy} updating.


\balance


\begin{spacing}{0.92}
\bibliographystyle{abbrv} 
\bibliography{IEEEtran,AOI-test}
\end{spacing}
\end{document}